# Microtubule defects influence kinesin-based transport in vitro

Winnie H. Liang,[1] Qiaochu Li,[1] K M R. Faysal,[1] Stephen J. King,[2] Ajay Gopinathan,[1] and Jing Xu[1, *]

[1]Department of Physics, University of California, Merced, CA 95343

[2]Burnett School of Biomedical Sciences, University of Central Florida, FL 32827

[*]**Correspondence:** Jing Xu (jxu8@ucmerced.edu)

## ABSTRACT

Microtubules are protein polymers that form "molecular highways" for long-range transport within living cells. Molecular motors actively step along microtubules to shuttle cellular materials between the nucleus and the cell periphery; this transport is critical for the survival and health of all eukaryotic cells. Structural defects in microtubules exist, but whether these defects impact molecular-motor based transport remains unknown. Here we report a new approach that allowed us to directly investigate the impact of such defects. Using a modified optical-trapping method, we examined the group function of a major molecular motor, conventional kinesin, when transporting cargos along individual microtubules. We found that microtubule defects influence kinesin-based transport in vitro. The effects depend on motor number: cargos driven by a few motors tended to unbind prematurely from the microtubule, whereas cargos driven by more motors tended to pause. To our knowledge, our study provides the first direct link between microtubule defects and kinesin function. The effects uncovered in our study may have physiological relevance in vivo.

## INTRODUCTION

Microtubules are biopolymers that self-assemble from tubulin dimers (1-5). During self-assembly, tubulin dimers stack longitudinally to form linear protofilaments, with multiple protofilaments associating laterally to form a hollow tubular structure, the microtubule (Fig. S1 *A* in the Supporting Material). Each microtubule is not necessarily perfect and can exhibit packing mistakes in the tubulin dimers (defects). The range of defects in microtubules include missing tubulin dimers (4, 5) and changes in the number of assembled protofilaments (1-3) (Fig. S1 *B*). These defects have been observed for microtubules in vitro (1-5) and in cell extracts (1).

Molecular motors such as kinesin rely on microtubules as molecular highways to drive mechanical transport in cells (6-10). This transport is critical for eukaryotic cell function and survival. Each individual kinesin typically tracks a single protofilament in each microtubule (11). Since microtubule defects include disruptions within individual microtubule protofilaments (Fig. S1 *B*), we hypothesized that these defects may influence kinesin-based transport.

A key experimental challenge for testing our hypothesis is that microtubule defects cannot be directly observed in current motility experiments using optical microscopes. Label-free imaging is not yet possible because the physical size of microtubule defects is ~1/20$^{th}$ below the optical resolution limit. There are also no known biomarkers for specific and non-invasive labeling/imaging of these structural defects, as the biochemical nature of the tubulin dimer within/surrounding these lattice defects is not yet understood.

To overcome these experimental challenges, we developed a single-microtubule assay to probe the effects of microtubule defects on kinesin-based transport in vitro. Since molecular motors typically work in small groups to transport materials in cells (6-10), we focused our investigations on cargo transport by more than one kinesin. To address the effects on transport by different motor numbers, we examined two regimes: one in which each bead was carried by a few kinesins and one in which each bead was carried by many motors. We found that microtubule defects influence kinesin-based transport in a manner that depends on the number of motors present on the cargo.

## MATERIALS AND METHODS

### Proteins and reagents

Kinesin and tubulin were purified from bovine brains as previously described (12, 13). Kinesin, which was microtubule-affinity purified and free of "dead" motors (12), was flash frozen in PMEE buffer (35 mM piperazine-N,N'-bis(2-ethanesulfonic acid) (PIPES), 5 mM MgSO$_4$, 1 mM EGTA, 0.5 mM EDTA, pH 6.8) supplemented with 45% glycerol and 1 mM dithiothreitol. Tubulin, which was free of microtubule-associated proteins (13), was flash frozen in PM buffer (100 mM PIPES, 1 mM MgSO$_4$, 2 mM EGTA, pH 6.9) supplemented with 45% glycerol and 1 mM dithiothreitol. Anti-tubulin antibody (T7816, clone SAP.4G5), poly-L-lysine (P8920), Pluronic F-127 (P2443), and chemicals (unless otherwise specified) were purchased from Sigma-Aldrich (St. Louis, MO, USA). Dimethyldichlorosilane solution (2% wt/vol, Repel-Silane ES) was purchased from GE Healthcare Bio-Sciences (Marlborough, MA, USA). Guanalyl-(α,β)-methylene diphosphate (GMPCPP) was purchased from Jena Biosciences (Jena, Germany).

**Microtubule preparation**

Microtubules were assembled in vitro and were free of microtubule-associated proteins. Two assembly conditions (taxol-stabilized and taxol-polymerized) were used to alter the frequency of defects in the assembled microtubules as previously described (2). Both taxol-stabilized and taxol-polymerized microtubules were kept at room temperature in a dark box and used within 8 days of preparation.

Taxol-stabilized microtubules were first assembled in the presence of GTP and then stabilized using taxol. Tubulin (40 µM) was supplemented with 0.5 mM GTP and incubated for 20 min at 37°C. The assembled microtubules were then diluted to 4 µM in PM buffer supplemented with 10 µM taxol, followed by a second incubation for 20 min at 37°C.

Taxol-polymerized microtubules were assembled in one step in the presence of taxol. Due to the increased microtubule-assembly rate in the presence of taxol (2), tubulin solution was diluted to 4 µM in PM buffer supplemented with 10 µM taxol and incubated for 30 min at 37°C for assembly as previously described (2).

A third assembly condition was used to alter the fraction of microtubules displaying supertwist (14, 15). Tubulin (4 µM) was supplemented with 1 mM GMPCPP and incubated for 3 h at 37°C (16). The assembled microtubules were diluted to 50 nM in PM buffer supplemented with 10 µM taxol and immediately introduced into flow cells for motility experiments. We refer to this third type of microtubule as "GMPCPP" microtubules.

**Flow cell preparation**

Motility investigations were carried out in flow cells in vitro. Flow cells were constructed by sandwiching a coverslip (22x40 mm, No. 1.5, Thermo Fisher Scientific Inc., Waltham, MA, USA) and a microscope slide (25x75 mm, Thermo Fisher Scientific Inc.) using double-sided Scotch tape. Both the microscope slide and the coverslip were biologically clean. The coverslip surface was either coated with poly-L-lysine or silanized to immobilize microtubules.

To immobilize microtubules through non-specific interaction with poly-L-lysine, the coverslip surface was plasma cleaned and then incubated with poly-L-lysine (0.00027% w/v in ethanol, 10 min). The coverslip was then oven dried (85°C, 10 min) prior to flow-cell construction. The microscope slide was not exposed to poly-L-lysine in order to minimize the presence of poly-L-lysine in the flow cell. Microtubules (taxol-stabilized or taxol-polymerized) were diluted to 50 nM in PMEE buffer (pH 7.2) supplemented with 1 mM GTP and 10 µM taxol and introduced into the flow cells. These microtubules underwent non-specific binding to the poly-L-lysine-treated coverslip surface. The flow cell was rinsed with wash buffer (11.7 mM PIPES, 1.6 mM MgSO$_4$, 0.3 mM EGTA, 0.12 mM EDTA, pH 7.2) supplemented with 1 mM GTP and 10 µM taxol and then blocked with 5.55 mg/mL casein solution in PMEE buffer supplemented with 1 mM GTP and 10 µM taxol. The resulting flow cell typically contained one or two isolated microtubules extending across the full width of our field of view. We used the same procedures to make flow cells with GMPCPP microtubules, except that we excluded GTP from the buffers. Unless otherwise indicated, experiments in the current study were carried out using polylysine-based immobilization.

To immobilize microtubules using anti-tubulin antibody, the coverslip surface was plasma cleaned and then treated with a dimethyldichlorosilane solution (2% wt/vol, 5 min). The coverslip was immersed in ethanol twice (5 min each), rinsed in Nanopure water three times, and air dried prior to flow-cell construction (17). The flow cell was incubated with 20 μg/mL anti-tubulin antibody in PEM80 buffer (80 mM PIPES, 4 mM $MgSO_4$, 1 mM EGTA, pH 6.9) for 5 min, and surface-blocked using 1% Pluronic F-127 in PEM80 buffer (18). Taxol-stabilized microtubules were diluted to 200 nM in PMEE buffer (pH 7.2) supplemented with 1 mM GTP and 10 μM taxol and introduced into the flow cells for 10 min. Excess microtubule was washed out using PEM80 buffer supplemented with 20 μM taxol and 10 mM dithiothreitol.

**Motor/bead preparation**

Carboxylated polystyrene beads (0.2 μm, Polysciences Inc., Warrington, PA, USA) were used to facilitate optical trap-based motility measurements. In experiments using polylysine-supported microtubules, kinesin was incubated with beads in motility buffer (67 mM PIPES, 50 mM $CH_3CO_2K$, 3 mM $MgSO_4$, 1 mM dithiothreitol, 0.84 mM EGTA, 10 μM taxol, pH 6.9) for 10 min at room temperature. In experiments using antibody-supported microtubules, kinesin was incubated with beads in motility buffer supplemented with 0.04% Pluronic F-127 (17). The motor/bead solution was then supplemented with an oxygen-scavenging solution (250 μg/mL glucose oxidase, 30 μg/mL catalase, 4.6 mg/mL glucose, Sigma-Aldrich) and 1 mM ATP prior to motility measurements.

We examined both few-motor transport and many-motor transport in the current study. To reach the few-motor range, we empirically tuned the kinesin-to-bead ratio such that the mean bead travel distance using measurements pooled over multiple microtubules was similar to that previously reported for transport by exactly two kinesins (19-22) (Fig. S2). To reach the many-motor range, we increased the motor-to-bead ratio by ~4-fold from the few-motor range.

For the few-motor system studied using polylysine-supported microtubules (Figs. 1 and S2-S4), 1.4 nM kinesin was incubated with a solution of $2.8 \times 10^6$ beads/μL. We reduced this motor-to-bead ratio to ~1.3 nM kinesin per solution of $2.8 \times 10^6$ beads/μL in Figure 2 in order to access a somewhat lower motor number in the few-motor range.

For the few-motor system studied using antibody-supported microtubules (Fig. 3), we found that the presence of 0.04% Pluronic F-127 in the motility buffer resulted in a somewhat lower level of motor/bead binding (data not shown). We therefore used higher concentrations of motors and beads during incubation while keeping the motor-to-bead ratio the same as that in Figures 1 and S2-S4. Specifically, 3.1 nM kineisn was incubated with a solution of $6.2 \times 10^6$ beads/mL for 10 min. The resulting motor/bead solution was diluted to $2.8 \times 10^6$ beads/mL to achieve the bead density suitable for optical trapping.

For the many-motor system studied using polylysine-supported microtubules (Figs. 4-6, S5 *A*, S8, S9 *A*, S11, and S12), 4.4 nM kinesin was incubated with $2.3 \times 10^6$ beads/μL.

For the many-motor system studied using antibody-supported microtubules (Figs. S5 *B*, S9 *B*, and S10), 8.8 nM kinesin was incubated with $4.6 \times 10^6$ beads/μL. The resulting motor/bead solution was diluted to $2.3 \times 10^6$ beads/μL for optical trapping.

**In vitro optical trapping**

A single-beam optical trap (23) was constructed and integrated with differential interference contrast imaging using an inverted microscope (Ti-E, Nikon, Melville, NY, USA). The optical trap (~20 mW at the laser output) was used to position individual kinesin-coated beads to a unique position on each microtubule under study. Throughout each experiment, we maintained both the optical trap and the microtubule in a static position and allowed Brownian fluctuation to bring the kinesin-coated beads to the optical trap. Upon observation of directed motion, the optical trap was manually shut off to allow bead motility without external load. We enhanced the Brownian motion of the kinesin-coated beads by using beads 0.2 μm in diameter (24), smaller than the usual 0.5-1 μm-diameter beads in optical trapping studies (23). The density of beads in our flow cell was empirically tuned to further enhance the probability that an individual bead is immobilized by the static optical trap, while minimizing the probability of multiple beads being trapped at the same time.

Bead trajectories were imaged at 250x magnification using differential interference contrast microscopy. Video data were recorded using a Giga-E camera (Basler SCA640-70GM, Basler AG, Ahrensburg, Germany) at 30 Hz. For each microtubule segment, 50-150 trajectories were measured, typically using different beads in the same flow cell.

**Data analysis**

Bead trajectories were particle-tracked to 10 nm resolution (1/3 pixel) using a template-matching algorithm as previously described (25). For each trajectory, travel distance was determined as the net displacement of the bead along the microtubule axis upon the bead's binding to and then detaching from a microtubule. The distribution of travel distances for each microtubule (or pooled using measurements from multiple microtubules) was fitted to a single exponential decay (26). To account for the time that elapsed during manual shutoff of the optical trap, only bead trajectories >0.3 μm were analyzed. Mean travel and the associated standard error for each distribution were determined as the fitted decay constant and associated uncertainty.

Unusual features in single-microtubule travel distributions were identified by comparing each distribution to its best-fitted value assuming the usual single-exponential decay function (26). An increase in counts was scored if the measurement was (a) greater than three times its best-fitted value and (b) at least 20% of the maximum measurement in the distribution. A decrease in counts was scored if the best-fit count value was (a) greater than three times the measured value and (b) at least 20% of the maximum measurement in the distribution. Note that our selection criteria for these unusual features were strict (in particular criterion (b) for both an increase and a decrease in counts) and did not include all substantial deviations between the measurements and the fitted values detected (for example, magenta arrows in Fig. 3 *B*).

Pausing was defined as the interruption in continuous bead motion along the microtubule axis. A pause event was scored when the instantaneous velocity of a bead (evaluated over 1 s) fell below a threshold of 0.15 μm/s (21). To account for thermal drift in our sample stage (up to 150 nm in amplitude over the typical 1 h duration of each experiment), we binned the on-axis position of each microtubule into discrete locations (180 nm bin width). Note that this coarse binning is implicit in the distribution of travel distances (650 nm bin, Fig. S2). For each microtubule, the mean number and standard deviation of trajectories pausing at each discrete location on a microtubule was calculated using all trajectories measured for that microtubule; a common pause location was identified when the number of pauses at a particular on-axis position was >4 standard deviations above the mean.

Pausing probability for each microtubule was determined as the fraction of trajectories that paused at least once on the microtubule. Standard error for pausing probability was calculated as the 68.3% confidence interval for a binomial distribution. The paired-sample t-test was used to determine the statistical significance of the difference in pausing probability between taxol-stabilized and taxol-polymerized microtubules.

Off-axis positions for each bead trajectory were mean removed and sign corrected such that a positive off-axis value corresponded to the left side of the microtubule axis when facing the direction of kinesin travel. The mean off-axis position for each bead trajectory was calculated as the midpoint between the minimum and maximum off-axis positions measured. The distribution of off-axis positions of beads (during pausing and during motion) was summed over all trajectories for each microtubule; the normalized distribution of off-axis positions for each microtubule was averaged over all microtubules. This approach limited potential bias toward individual microtubules.

The rank-sum test was used to determine the statistical difference between two distributions of travel measurements. A one-way analysis of variance was used to determine the statistical difference between multiple distributions of travel measurements.

## RESULTS

**A single-microtubule assay to probe the effects of microtubule defects on cargo transport**

Here, we developed a simple assay to measure multiple cargo trajectories along the same microtubule segment (Fig. 1 *A*). Multiple measurements are necessary to sample defect sites on a microtubule surface: since a defect is small compared to the overall microtubule surface available for cargo transport, the probability that a particular cargo trajectory will encounter a defect site is likely to be small. Using a single-beam optical trap, we defined and maintained a unique interaction site for cargos on each microtubule (Fig. 1 *A*). This approach enabled us to repeatedly survey the same microtubule segment, thereby increasing the net number of trajectories encountering and being influenced by microtubule defects. We anticipated that these trajectories would provide a direct, functional readout on the hypothesized impact of microtubule defects on motor function. Since this approach focused on transport along individual microtubules, we refer to it as the "single-microtubule assay."

**Common unbinding sites on microtubules for transport by a few motors**

We first investigated the potential impact of microtubule defects on beads transported by a few kinesins (Fig. 1). We empirically tuned the motor-to-bead ratio such that the resulting mean bead travel distance (pooled over multiple microtubules) was within the range previously reported for transport by exactly two kinesins (assembled using DNA/protein-based structures) (19-22) (Fig.

S2). We refer to this range as "few-motor transport" because the number of motors decorating each bead is Poisson-distributed rather than well-defined in bead-based studies (26, 27) and in vesicle transport in vivo (6, 9). To capture the key characteristics of microtubule assembly in cells (28), we utilized microtubules assembled in the presence of GTP. To halt the dynamic disassembly of microtubules during our motility studies in vitro, we stabilized these assembled microtubules via taxol (29), and refer to them here as "taxol-stabilized microtubules."

We observed significant variations in single-microtubule travel distributions measured under otherwise identical conditions (Fig. 1). A total of 33 single-microtubule travel distributions were measured, corresponding to ~70 trajectories along each microtubule. A one-way analysis of variance revealed significant differences among the 33 single-microtubule travel distributions, $F(32, 2434) = 9.16$, $P < 0.001$. To control for the possibility that these travel variations reflect experimental variations between kinesin/bead preparations, we focused on the subset of measurements for different microtubules in the same flow cell (Figs. 1 *B* and S3). Since the same population of kinesin-decorated beads was present in each flow cell, the only variable in our experiments was the microtubule along which the beads traveled. For these side-by-side measurements, we again observed significant travel differences between microtubules for 3/8 triplet sets (Figs. 1 *B* and S3).

What is responsible for these travel variations? To address this question, we examined each of the 33 single-microtubule travel distributions. For 6/33 microtubules (18%), we observed unusual features indicating common unbinding sites on each microtubule (Figs. 1 *D* and S4). For example, rather than the distribution being well approximated by the usual single exponential decay (26), we observed 11-fold more counts than expected for the travel distance of ~5.7 μm (magenta arrow, Fig. 1 *D*), and 6-fold fewer counts than expected for the travel distance of ~2.5 μm (orange arrow, Fig. 1 *D*). These substantial deviations reflect common unbinding sites on the microtubule: since each trajectory shared the same starting position on a microtubule (red line, Fig. 1 *C*), trajectories with the same travel distance must unbind at the same location along the microtubule (magenta arrow in Fig. 1 *D* and magenta line in Fig. 1 *C*). Note that when the common unbinding event is located near the initial travel position, the increase in counts biases the fitting toward a longer decay constant, resulting in an apparent reduction in counts in the travel distribution (orange arrow, Fig. 1 *D*). The locations of these unusual features differed between microtubules without apparent periodicity (Fig. S4 *A*); the magnitudes of these unusual features diminished in travel distributions pooled from measurements using multiple microtubules (Fig. S2).

Perhaps strikingly, despite the presence of more than one motor per cargo, the mean travel distance for 3/33 microtubules (9.1%) was substantially smaller than the average travel distance for a single kinesin (Figs. 1 *B* and S4 *B*). For example, beads traveled $0.63 \pm 0.05$ μm (mean ± standard error, $n = 64$) along MT4, corresponding to a 37% reduction below single-kinesin travel (26) (Fig. 1 *B*). This reduction in mean travel distance cannot be simply explained by experimental variation in the kinesin-to-bead ratio or by a loss of motor activity, since the same population of beads gave rise to significantly longer mean travel distances for the two microtubules measured at later time points in the same flow cell (MT5 and MT6 in Fig. 1 *B*). Instead, it is consistent with local clustering of one or more unbinding sites near the initial travel position on the microtubule.

Together, these data support our hypothesis that microtubule defects influence cargo transport, for example by prompting kinesins to dissociate prematurely from the microtubule.

**No significant effect of microtubule supertwist on cargo travel/unbinding**

Next, we sought to control for the possibility that common unbinding sites were artifacts of microtubule supertwist (Fig. 2), the rotation of individual protofilaments along the microtubule axis. Since each kinesin typically tracks a single protofilament during transport (11), it is formally possible that the cargo travel may unbind at the interface between rotating protofilaments and the coverslip surface that supports the microtubule.

We tested this possibility by increasing the probability of each cargo encountering a microtubule/coverslip interface (Fig. 2 *A*). Previous studies reported that ~40% of taxol-stabilized microtubules (which we used thus far) have supertwist (2), whereas ~95% of GMPCPP microtubules (Materials and Methods) have supertwist (14, 15). We therefore generated GMPCPP microtubules for comparison with taxol-stabilized microtubules (16). We used a single population of tubulin dimers to keep the biochemical makeup of these two types of microtubules constant. We also used the same kinesin/cargo preparation (Fig. 2 *A*) and altered the measurement order of the microtubule types for each set of pairwise comparisons. A total of six microtubule pairs were tested.

To isolate the effect of supertwist and to minimize the impact of microtubule defects on these comparisons, we employed the standard multiple-microtubule assay (26) (Fig. 2 *B*). We measured one and only one cargo trajectory for each microtubule, and sampled ~200 microtubules to obtain the average travel distance along each microtubule type. The probability that each trajectory encounters the defect on a microtubule is low, and the locations of defects likely differ between microtubules. Thus, the key difference between the two microtubule types is the probability of a microtubule displaying supertwist (40% (2) vs. 95% (14, 15)).

We did not detect any significant difference in cargo travel between these two types of microtubules (Fig. 2 *C*). Note that to amplify the sensitivity of cargo travel to this potential surface effect, we used fewer motors per cargo here than in the experiments in Figure S2. Cargo transport remained in the few-motor range, with a mean travel distance longer than that resulting from transport by a single kinesin (>1.1 μm in Fig. 2 *C* vs. ~1 μm for the native bovine kinesin used in this study (26)). Our data demonstrate that microtubule supertwist does not contribute significantly to the travel variations and common unbinding events in our few-kinesin system.

**Cargo unbinding occurs independently of the microtubule-immobilization method**

To further investigate the possibility of surface effects, we used anti-tubulin antibody to elevate microtubules above the coverslip surface (Fig. 3 *A*), in addition to the polylysine-supported microtubules investigated thus far. We also employed a different surface-blocking agent (Pluronic F-127 for the antibody-based immobilization vs. casein for polylysine-based immobilization) to reduce non-specific interactions between motor-decorated beads and the coverslip surface. We measured few-motor travel along different taxol-stabilized microtubules using the same population of motor/bead complex in a single flow cell (Fig. 3 *B*).

We detected significant differences in travel distance between antibody-supported microtubules (asterisks, Fig. 3 *B*). For example, the rank-sum test returned a P-value of 0.0024 for travel along $MT_{antibody}2$ versus $MT_{antibody}3$ (asterisks, Fig. 3 *B*). One-way analysis of variance also revealed significant differences among the three single-microtubule travel distributions ($F(2, 187) = 7.57$, $P < 0.001$).

Data from antibody-supported microtubules also substantially deviated from the typical single exponential decay (Fig. 3 *B*). For example, for $MT_{antibody}2$, we observed 8.5-fold more counts than expected for the travel distance of ~5.5 μm (magenta arrow, Fig. 3 *B*). For $MT_{antibody}3$, we detected 3.4-fold fewer counts than the fitted value for the travel distance of ~2.3 μm (orange arrow, Fig. 3 *B*), as well as >11-fold more counts than the fitted values for travel distances of 3.9-5.5 μm (magenta arrows, Fig. 3 *B*). Taken together, these data demonstrate that cargo unbinding during few-motor transport is not specific to any particular microtubule immobilization method.

**Common pause locations on microtubules for many-motor transport**

Next, we increased the motor-to-bead ratio by ~4-fold to reach the many-motor transport range (Materials and Methods). The resulting cargo travel distance increased significantly versus that of the few-motor system (>20 μm in Fig. 4 vs. <2 μm in Fig. S2), suggesting that unbinding events were substantially suppressed by the increase in motor number. We hypothesized that the extended travel in this many-kinesin system (Fig. 4 *A-B*) would expand the range of effects detectable by our assay as well as increase the number of defects encountered by each trajectory. This many-motor range may also shed light on the long-range transport of large cargos in vivo, such as the movement of mitochondria in neuronal processes and nuclear migration (30-32). For consistency, we continued to use taxol-stabilized microtubules and polylysine-based immobilization as we did for experiments in Figure 1.

We found that the probability of cargo pausing during continuous transport increased from $2.7 \pm 0.4\%$ for few-kinesin transport to $13.2 \pm 1.2\%$ for many-kinesin transport (arithmetic mean ± standard error; $n = 33$ and 42 microtubules, respectively). This observation is consistent with previous findings (using multiple-microtubule assays) that pausing in kinesin-based transport occurs minimally for single kinesins (33-35) and more frequently for multiple motors (17, 21).

Interestingly, our single-microtubule study revealed common pause locations for 9/42 microtubules (Figs. 4 *A-B* and S5 *A*). For example, seven trajectories paused at the same site on MT34 (~8 μm, Fig. 4 *A*), and 12 trajectories paused at a single site on MT35 (~16.4 μm, Fig. 4 *B*). Both of these pause counts were >4 standard deviations above the mean for each microtubule. The positions of these common pauses were relative to the position of our optical trap, and did not correspond to any underlying rotational pitch of the microtubule. The locations of these common pause sites also differed between microtubules (Fig. S5 *A*). Together, these findings are consistent with our hypothesis that microtubule defects influence kinesin-based transport, for example by triggering cargo pausing in many-kinesin transport.

**Cargo trajectories during pausing reflect force-based interactions between motors**

For all pauses, we observed that the cargo velocity gradually slowed to zero while entering a pause, and recovered to normal levels after the pause (~0.8 μm/s, Fig. 4). The initial slowdown indicates that the motors transporting these cargos experienced substantial force opposing their motion, as kinesins respond to forces opposing the direction of their travel by slowing (27, 36, 37). Since we turned off the optical trap upon observation of directed bead motion, the source of opposing force must be internal to the team of motors transporting the same cargo: a subset of motors on the cargo paused on the microtubule and hindered other motors from pulling the cargo forward. The abrupt velocity recovery when exiting a pause is consistent with dissociation of the paused motors, which allows the motors to move forward without load at normal velocity.

We observed two types of cargo trajectories during pausing: static (Fig. 4 *C*) and dynamic (Fig. 4 *D-E*). During static pausing, which occurred with all 42 microtubules examined, cargo velocity remained unchanged at zero (Fig. 4 *C*). During dynamic pausing, which was observed for 10/42 microtubules, the cargo underwent substantial backward and forward movements while its net position remained approximately constant (backward and forward arrows, Fig. 4 *D-E*). Cargo velocity was 3-fold faster during backward movements than during forward movements (Fig. 4 *D-E*).

Backward movements are surprising, since only one type of motor was present to drive cargo transport. While it is formally possible that a bead dissociates from the microtubule briefly and rebinds, or that the bead rotates backward, allowing other motors to bind, neither scenario explains the asymmetry in cargo velocity during backward and forward movements (2.0 μm/s vs. 0.66 μm/s, Fig. 4 *D-E*). Instead, fast backward movements are consistent with cargo "flopping" back to the location of paused kinesins (17, 38) when the leading motor stochastically unbinds from the microtubule (Fig. S6). Slow forward movements are consistent with the rebinding of detached motors driving the cargo forward and being hindered by the paused motor(s) lagging behind. Occasionally, we detected somewhat slower backward movement (backward movement at ~17.5 s, Fig. 4 *E*) that may be due to successive unbinding of a few leading motors. The range of these backward movements is consistent with the flopping mechanism, given kinesin's contour length (39, 40) and our bead size (Fig. S6).

**Surface effects are not responsible for cargo pausing in many-motor transport**

Are surface effects responsible for pausing in many-motor transport? To address this possibility, we examined the off-axis position of cargos during pauses and during motion (Fig. 5). If a surface effect (for example microtubule supertwist) is the main cause of pausing, then beads should pause preferentially at the interface between the microtubule and its coverslip support. Extended travel in the many-kinesin system allowed us to determine the distribution of the off-axis positions for each bead trajectory (Fig. 5 *A*). The range of off-axis position (~200 nm) is reasonable considering kinesin's contour length (39, 40) and our bead size (Fig. S7). Note that this approach was not possible for the few-kinesin system because travel distances were too short to fully map the off-axis range of a microtubule. Some of the microtubules in our experiments underwent thermal motion relative to the coverslip support (Fig. S8 and Movie S1). Thus, to examine motor-based bead motion independent of thermal motion of the microtubule, we focused this analysis on the subset of 34 microtubules that were well anchored to the coverslip throughout the experiment.

We did not observe any tendency of beads to pause at the microtubule/coverslip interface (Fig. 5 *B-C*). For each microtubule, pauses did not occur at the same off-axis position (Fig. 5 *B*). For example, despite sharing a common on-axis location, pauses 1-3 differed significantly in their off-axis position (Fig. 5 *B*). Note that the constrained off-axis diffusion during pausing in Figure 5 *A* was not a general observation (Fig. S9 and the supporting text). When we pooled measurements from 34 microtubules, the distribution of off-axis positions during pausing was in excellent agreement with that during motion (red vs. black lines, Fig. 5 *C*). Both distributions were well described by a normal distribution ($R_{adj}^2$ = 0.95 and 0.99, respectively) that is centered about the mean off-axis position of the microtubule (0.3 ± 3.6 nm and 6.3 ± 1.6 nm, respectively). These analyses demonstrate that cargos did not preferentially pause at the microtubule/coverslip interface, indicating that surface effects are not the main cause of pausing in our many-kinesin system.

We also carried out control experiments in which we elevated the microtubule above the coverslip surface (Fig. S5 *B*). We found that our findings on cargo pausing thus far were independent of the microtubule-immobilization method (Figs. S5, S9 and S10). Using antibody-supported microtubules, we again detected common pause sites along the microtubule axis (Fig. S5 *B*) and observed both static and dynamic cargo trajectories during pausing (Fig. S10). We also observed instances of constrained off-axis diffusion during pausing (Fig. S9 *B*). Again, this constrained off-axis diffusion during pausing was not a general observation (Fig. S9 *B*).

**Pausing probability increases for microtubules with higher defect frequency**

Next, we examined whether pausing probability was directly affected by the frequency of defects in microtubules. To isolate the effect of defect frequency on pausing, we used identical flow cells (polylysine-based) to harbor microtubules with different defect frequencies.

We generated taxol-polymerized microtubules for comparison with taxol-stabilized microtubules (Materials and Methods). Arnal and Wade previously reported that the frequency of defects in a microtubule varies to some extent with microtubule-assembly conditions (2). Specifically, the number of transitions in protofilament number within a microtubule is twice as high for taxol-polymerized microtubule as for the taxol-stabilized microtubules that we have described thus far (2). We used a single population of tubulin dimers to keep the biochemical makeup of these two types of microtubules constant. We also used the same kinesin/cargo preparation (Fig. 6 *A*) and altered the measurement order of the microtubule types for each set of pairwise comparisons. Microtubule immobilization was achieved using the polylysine-based method. A total of 14 microtubule pairs were tested. Consistent with our observations with taxol-stabilized microtubules (Fig. 5), cargo pausing on individual taxol-polymerized microtubules did not occur at the same off-axis position (Fig. S11 *A-B*), and surface effects were not the main cause of cargo pausing (Fig. S11 *C*).

Importantly, we observed a significantly higher probability of pausing on microtubules with higher defect frequency (taxol-polymerized microtubules) for 4/14 pairwise comparisons (microtubule pairs 1, 8, 10, and 11, Fig. 6 *B*). We did not detect any instances in which a significantly higher pausing probability occurred for microtubules with lower defect frequency (taxol-stabilized microtubules). Overall, the paired-sample t-test indicated that pausing probabilities differed significantly between these two microtubule types ($P$ = 0.028, Fig. 6 *B*).

The mean pausing probability was 1.6 ± 0.3-fold larger for microtubules with a higher defect frequency (Fig. 6 *C*). When we calculated the ratio of pausing probability for each pairwise comparison, we uncovered an average 2.2 ± 0.6 higher likelihood of a trajectory pausing on taxol-polymerized microtubules (more defects) than on taxol-stabilized microtubules (fewer defects). These data indicate that microtubule defects were the dominant factor underlying cargo pausing in the many-motor system.

We observed similar trends in the number of pause locations along each microtubule axis (Fig. S12). For 4/14 comparisons with significantly higher pausing probability for microtubules with more defects, the number of pause locations was more than 2-fold larger (microtubule pairs 1, 8, 10, and 11, Fig. S12 *A*). We observed only one instance in which the number of pause locations was substantially (2-fold) higher for microtubules with lower defect frequency (microtubule pair 5, Fig. S12 *A*). Overall, the paired-sample t-test demonstrated that the number of pause locations was substantially different between the two microtubule types ($P = 0.055$, Fig. S12 *A*). The mean number of pause locations was 1.5 ± 0.3-fold larger for microtubules with higher defect levels (Fig. S12 *B*). When we calculated the ratio of pause locations for each pairwise comparison, we detected an average of 2.0 ± 0.4 more pause locations on taxol-polymerized microtubules than on taxol-stabilized microtubules. These data are consistent with the previous finding that the frequency of defects in taxol-polymerized microtubules is twice that in taxol-stabilized microtubules (2).

Taken together, our data demonstrate that cargo pausing is directly influenced by microtubule defects. As the number of defects in the microtubule increases, the probability that a cargo will pause along that microtubule also increases.

**DISCUSSION**

Here we used a single-microtubule assay to probe the functional importance of microtubule defects on kinesin-based transport in vitro. This approach differs from standard multiple-microtubule assays in that it specifies the microtubule for transport, thus yielding information about cargo transport as well as the "road condition" of the microtubule. Our data indicate that microtubule defects influence kinesin-based transport in vitro, prompting cargos to unbind from the microtubule (Fig. 1) or to pause during continuous motion (Fig. 4). Importantly, these effects were independent of the microtubule immobilization method; we observed cargo unbinding and pausing for both polylysine-supported microtubules (Figs. 1 and 4) and antibody-supported microtubules (Figs. 3, S5 *B*, S9 *B*, and S10).

We did not detect a significant role of surface effects on cargo unbinding (Figs. 2 and 3) or pausing (Figs. 5, S5, S9, and S10). We also observed significantly more cargo pausing on microtubules with more defects (Fig. 6), indicating that the main factor determining cargo pausing is microtubule defects, not experimental artifacts. Note that measurements from our few-motor study in Figure 2 do not conflict with a previous report of a somewhat shorter single-kinesin travel distance along GMPCPP microtubules than along taxol-stabilized microtubules (41). Kinesin has a 3.7-fold higher binding affinity for GMPCPP microtubules than for taxol-stabilized microtubules (42, 43). Travel distance in the few-motor system is sensitive to both the binding and unbinding events of individual motors with the microtubule (44), and need not reflect the trend of single-motor travel.

How do microtubule defects cause cargo unbinding and pausing? Microtubule defects include missing tubulin dimers (4, 5) and changes in the number of protofilaments within a microtubule (1, 2). Previous studies demonstrated that kinesins require successive tubulin dimers to sustain transport (45). We thus propose that the main effect of a missing tubulin dimer is to prompt kinesin to unbind from the microtubule. Kinesins also tend to pause in crowded environments (17, 46, 47). A change in protofilament number leads to the merging of two or more protofilaments into one "lane" when the transition occurs in the direction of kinesin transport. Kinesins that track merging protofilaments likely crowd into a "traffic jam" at the merging site. We therefore propose that the main effect of protofilament merging is to cause a subset of motors to pause at the merging site, hindering cargo transport.

Our model accounts for our finding that the effects of microtubule defects strongly depend on the number of motors available for transport (Figs. 1 and 4). For the few-kinesin system, the number of motors available to create a traffic jam is limited, but each cargo is more sensitive to the unbinding of individual motors. The effect of individual motor unbinding is suppressed in the many-kinesin system, but more motors track the merging protofilaments and can contribute to a traffic jam. We are working on future methods to generate microtubules that preferentially express each of these two defect types.

The effects uncovered in our study may have physiological relevance in vivo. These local changes in transport (shorter travel distance or slower velocity) may have downstream consequences for cellular functions that rely on proper cargo transport. The magnitude of this impact remains unknown, since various microtubule-associated proteins decorate microtubules and may obscure these defects from cargo transport in vivo. However, there is evidence that the structure of microtubules is tightly regulated in cells. The microtubule-severing protein katanin targets and removes microtubule defects (48, 49), and the microtubule-associated proteins doublecortin and EB1 preferentially stabilize microtubules with 13 protofilaments (3, 50, 51). Our study raises the intriguing possibility that an important function of these machineries may be to maintain "road conditions" for cargo transport in vivo.

## CONCLUSION

In the current study, we developed a simple assay to examine kinesin-based transport along individual microtubules. We observed common unbinding sites on microtubules for few-motor transport as well as common pausing locations for the many-kinesin system. The trajectories of cargos during pausing reflected force-based interactions between paused and moving motors. Our control studies demonstrated that these surprising new effects were not specific to any particular microtubule-immobilization method. Few-kinesin travel was independent of the fraction of the microtubules displaying microtubule supertwist. We did not detect preferential pausing at the microtubule/coverslip interface. Instead, our data demonstrate that the probability of pausing in the many-kinesin system is directly tuned by the frequency of microtubule defects. Taken together, our study provides, to our knowledge, the first direct link between microtubule defects and kinesin function.

## SUPPORTING MATERIAL

Twelve supporting figures and one supporting movie are attached.

## AUTHOR CONTRIBUTIONS

JX designed the study. JX and QL performed the experiments. WHL, JX, QL, KMRF, and AG analyzed the data. SJK purified proteins. JX, SJK, and AG wrote the paper.

## ACKNOWLEDGMENTS

We thank Ahmet Yildiz, John A Hammer, Keir C Neuman, Steven P Gross, Stefan Diez, Felix Ruhnow, Erik Schäffer, Luke M Rice, and Jennifer L Ross for helpful discussions. We thank Kasimira Stanhope, Amanda J Tan, and Michael W Gramlich of the Ross lab for help on the antibody-based microtubule immobilization method. We thank Tiffany J Vora for helpful comments on the manuscript. We thank the editor and the reviewers for the thoughtful suggestions.

This work was supported by the UC Merced Senate Committee on Research (to JX), the National Institutes of Health (NS048501 to SJK), the National Science Foundation (PHY-1066293 and EF-1038697 to AG), and the James S McDonnell Foundation (to AG). Portions of this work were carried out by JX and AG at the Aspen Center for Physics, supported in part by NSF Grant PHY-1066293.

## SUPPORTING CITATIONS

References (52-54) appear in the Supporting Material.

**FIGURE LEGENDS**

FIGURE 1     Single-microtubule (MT) measurements of cargo travel for few-kinesin transport. (*A*) Schematic of our single-MT assay (not to scale). An optical trap directs kinesin-coated beads to a unique position on a MT. (*B*) Mean travel distances of beads measured for individual MTs. MT1-3, 4-6, and 7-9 represent sets of three MTs; each set was measured in the same flow cell. Error bars, standard error. Asterisks, significant differences in cargo travel distance between MT pairs ($P < 0.02$, rank-sum test). Corresponding single-MT travel distributions are shown in Figure S3. (*C*) Example single-MT trajectories sharing the same initial travel position (red dashed line) on a single MT. Each trajectory represents a different bead trapped from the same bead population in the flow cell; the trajectories are offset with regard to their relative timing (x axis) in order to facilitate comparison. *n*, total number of trajectories measured for this MT. Magenta dashed line, MT position at which several beads disengage from transport. (*D*) Single-MT travel distribution corresponding to trajectories in (*C*). Blue line, best fit to a single exponential decay. Mean travel distance ($d \pm$ standard error), goodness of fit ($R_{adj}^2$), and sample size (*n* trajectories) are indicated. Arrows, deviations from best fit (magenta, more counts; orange, fewer counts; see Materials and Methods).

FIGURE 2     Pairwise comparisons of travel distance along two types of microtubule that differed in their respective likelihood of displaying supertwist (95% and 40%). (*A-B*) Experimental schematic (not to scale). We used the same kinesin/bead mixture to contrast transport between the two types of microtubules (*A*). We used the standard multiple-microtubule

assay to minimize the influence of microtubule defects (*B*). (*C*) Distribution of few-kinesin travel along each microtubule type. Hatched bars at ~9 µm indicate travel distances that exceeded our field of view. Solid line, best fit to a single exponential decay. Mean travel distance ($d$ ± standard error) and sample size (*n* trajectories) are indicated. These distributions do not differ significantly from each other ($P = 0.36$, rank-sum test).

FIGURE 3    Measurements of cargo travel along antibody-immobilized microtubules (MTs). (*A*) Experimental schematic (not to scale). Anti-tubulin antibody was used to elevate the MT above the coverslip surface and Pluronic F-127 was used to reduce non-specific interactions between motor/bead complexes and the coverslip. (*B*) Single-MT travel distributions measured for three MTs in the same flow cell. Asterisks, significant differences in cargo travel distance between MT pairs ($P < 0.02$, rank-sum test). Blue line, best fit to a single exponential decay. Mean travel distance ($d$ ± standard error) and sample size (*n* trajectories) are indicated. Hatched bars at 12 µm indicate cumulative counts of travel distance that exceeded our field of view. Arrows, substantial deviations between measurements and best-fitted values.

FIGURE 4    Single-microtubule (MT) measurements of cargo pausing during many-kinesin transport. (*A-B*) Example trajectories (left) and the corresponding distribution of pauses along the MT axis (right) measured for two MTs. Each trajectory represents a different bead trapped from the same bead population in a single flow cell; the trajectories are offset with regard to their relative timing (x axis) in order to facilitate comparison. Red asterisks indicate common pause locations (>4 standard deviations above the mean number of pauses for that MT). (*C-E*) Example trajectories exhibiting static (*C*) and dynamic (*D-E*) pausing. Blue arrows indicate the direction of cargo travel. Mean cargo velocity (± standard error) during dynamic pausing (*D-E*) was 2.0 ± 0.4 µm/s ($n = 11$) during backward movement and 0.66 ± 0.16 µm/s ($n = 11$) during forward movement.

FIGURE 5    Distributions of off-axis positions of beads during pausing and during motion. (*A*) Example off-axis (top) and on-axis (bottom) positions for one bead trajectory. Vertical dash-dot lines indicate pausing. (*B*) Example distributions for six trajectories along the same microtubule. The distribution during cargo motion represents averages of all six trajectories; error bars, standard error. Distributions during cargo pausing were not averaged and represent individual trajectories. Pauses 1-3 shared the same on-axis location on the microtubule. Pause 6 corresponds to the off-axis trajectory shown in (*A*). (*C*) Normalized distributions averaged over 34 microtubules (MTs). Error bars, standard error.

FIGURE 6    Probability of cargo pausing on microtubules with different defect frequencies. (*A*) Schematic of experimental setup (not to scale). A single population of kinesin-coated beads was introduced into two flow cells containing taxol-stabilized microtubules (blue) or taxol-polymerized microtubules (orange). Asterisks illustrate the relative defect frequencies as previously reported (2). (*B*) Probability of a trajectory pausing on each microtubule. Error bars, standard error. (*C*) Distributions of pausing probability measured for each microtubule type. Mean pausing probability (± standard error) and sample size (*n* microtubules) are indicated.

# SUPPORTING MATERIAL

## Microtubule defects influence kinesin-based transport in vitro

Winnie H. Liang,[1] Qiaochu Li,[1] K M R. Faysal,[1] Stephen J. King,[2] Ajay Gopinathan,[1] and Jing Xu[1, *]

[1]Department of Physics, University of California, Merced, CA 95343

[2]Burnett School of Biomedical Sciences, University of Central Florida, FL 32827

[*]**Correspondence:** Jing Xu (jxu8@ucmerced.edu)

**RUNNING TITLE:** Microtubule defects impact kinesin function

## TABLE OF CONTENTS

## SUPPORTING TEXT

**Off-axis diffusion of beads may be constrained during pausing**

Off-axis diffusion of beads depends non-trivially on both the number of motors linking the bead to the microtubule, and the arrangement of bound motors across different microtubule protofilaments (off-axis distance between motors) (1, 2). Thus, constrained off-axis diffusion may indicate that there are many motors linking the bead to the microtubule, and/or reflect a large off-axis distance between bound motors.

We observed instances in which off-axis diffusion of beads was constrained during pausing (at ~10.5 s in Fig. S9 *A* i and at ~24 s in Fig. S9 *B* iii). Paused motors likely have longer association times with the microtubule (or lower unbinding rates) than moving motors (3). We therefore suspect that a microtubule defect may increase the duration of the particular binding arrangement of motors linking the cargo and the microtubule, as well as increase the instantaneous number of bound motors during cargo pausing. Both of these possibilities may decrease off-axis diffusion during pausing, in particular when the bound motors occupy a large off-axis distance across microtubule protofilaments. We previously detected a significant reduction in the off-axis diffusion of beads during motion when we lowered the unbinding rate of individual motors using a limiting ATP concentration (2).

However, the constrained off-axis diffusion during pausing was not a general observation in the current study (for example, pauses in Fig. S9 *A* ii and in Fig. S9 *B* iv, and at ~4 s in Fig. S9 *B* iii). The relatively un-constrained off-axis diffusion in these pauses likely reflects a small off-axis distance between the paused and the moving motors, and/or a low number of motors (paused or moving) linking the bead to the microtubule surface.

**Off-axis movement of beads is not directly coupled to pauses**

Off-axis movement was not a general observation for cargo pausing (for example, Fig. S9 *A* ii). Pausing and off-axis movement of a bead are not fundamentally related. In order for a bead to undergo off-axis movement, the number of motors linking the bead to the microtubule must change. Although changing the number of bound motors changes the number of motors available for pausing, it does not guarantee the presence or absence of pausing. Thus, we believe that there is no fundamental difference between trajectories demonstrating off-axis movement after pausing and trajectories lacking off-axis movement (for example, Fig. S9 *A* i vs. ii); both modalities rely on stochastic dissociation of paused motor(s), which allows the bead to resume motion. Similarly, we believe that there is no fundamental difference between trajectories demonstrating off-axis movement before/during a pause and trajectories lacking such off-axis movement. Such off-axis movement indicates only a change in the number of motors bound to the microtubule, but does not provide additional information on the nature of this change (change in motor number or pausing state of the motor).

**SUPPORTING FIGURES**

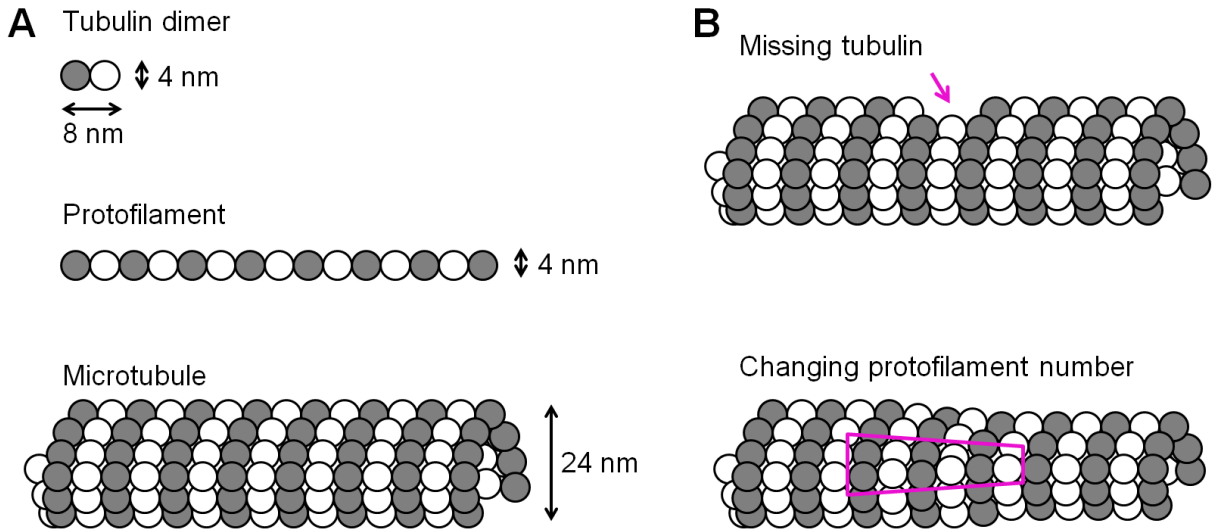

**Figure S1.** Schematic of microtubule self-assembly (*A*) and the range of microtubule defects reported by previous studies (4-7) (*B*). Illustrations are not to scale. (*A*) Microtubules are tubular structures formed via hierarchical self-assembly of tubulin dimers into protofilaments, which then associate to form a hollow tube (4-7). Tubulin dimers are heterodimers composed of α and β tubulin monomers (~4 nm in diameter), as indicated by grey and white spheres. (*B*) Defects in the microtubule structure were previously uncovered by cryoelectron microscopy (4-6) and scanning force microscopy (7). These defects include missing tubulin dimers (top) and changes in the numbers of protofilaments (bottom). The biochemical nature of these defects is not clear. The physical size of these defects is ~1/20[th] below the optical resolution limit. Direct visualization of these defects during motility assays is not currently possible and is outside of the scope of this study.

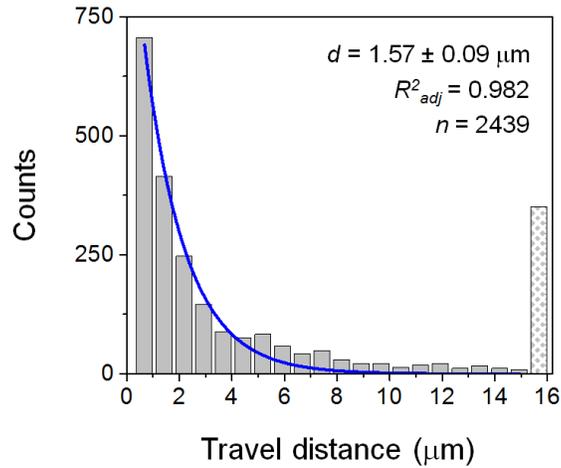

**Figure S2.** Distribution of travel distances measured for 33 taxol-stabilized microtubules (~70 trajectories for each microtubule). Kinesin (1.4 nM) was incubated with a solution of $2.8 \times 10^6$ beads/µL. Blue line, best fit to a single exponential decay. The shaded bar at 15.7 µm indicates the cumulative counts of travel distances that exceeded our field of view. Mean travel distance ($d$ ± standard error), goodness of fit ($R_{adj}^2$), and sample size ($n$ trajectories) are indicated. The mean travel distance is within the range previously reported for transport by exactly two kinesins (assembled using DNA/protein-based structures) (8-11). We used this kinesin-to-bead ratio to study bead transport by a few kinesins ("few-motor system").

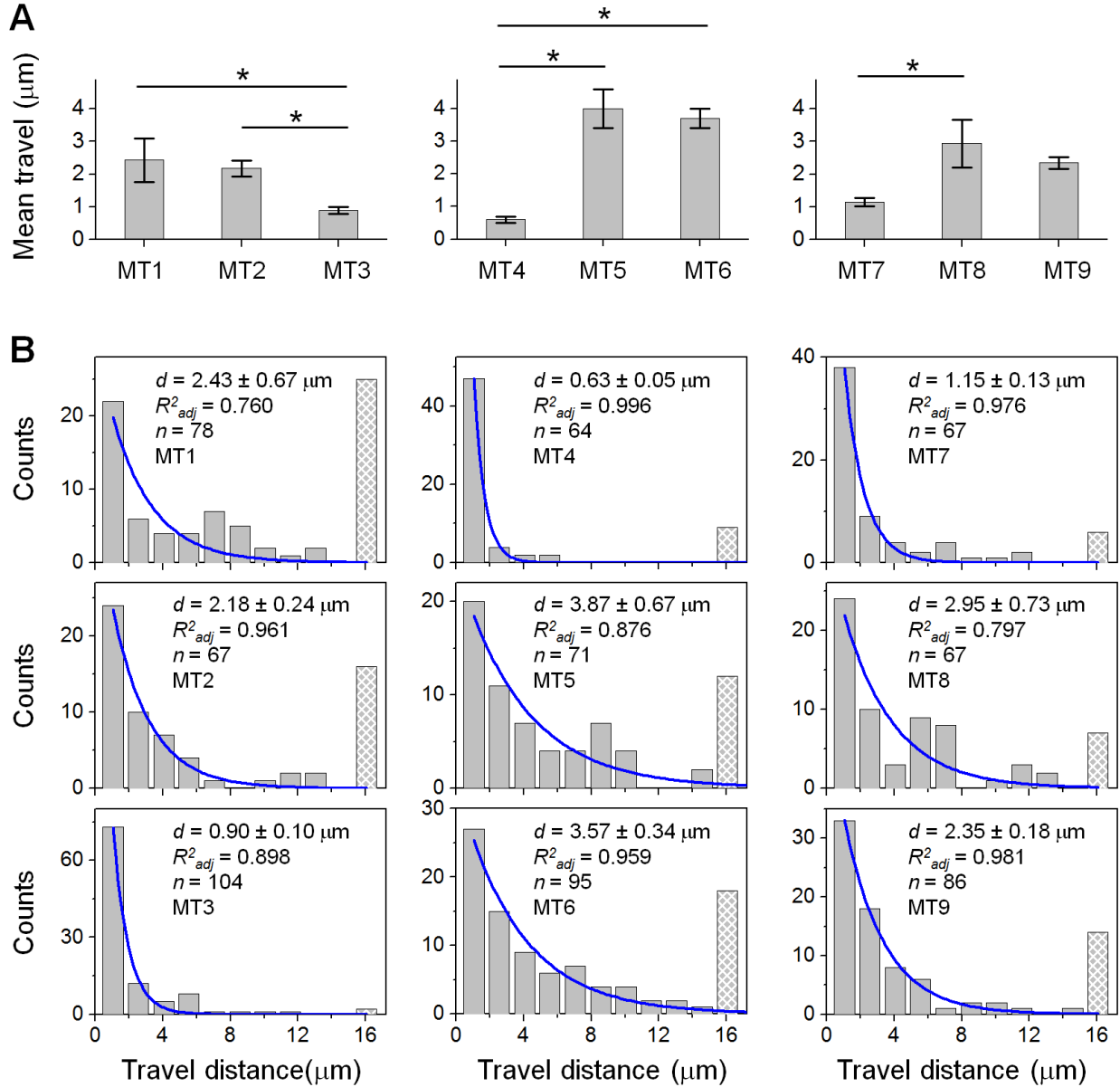

**Figure S3.** Measurements of few-motor travel distance along different taxol-stabilized microtubules (MTs) in the same flow cell, corresponding to data shown in Figure 1 *B* in the main text. We observed significant differences in bead travel for 3/8 triplet comparisons (37.5%). The kinesin-to-bead ratio was kept constant at 1.4 nM kinesins: $2.8 \times 10^6$ beads/μL (few-motor range, Fig. S2). (*A*) Mean travel distances for three sets of side-by-side comparisons. Each set used a single population of kinesin-coated beads in the same flow cell. Error bars, standard error. Asterisks, statistically significant differences in mean travel distance between MT pairs ($P < 0.02$, rank-sum test). (*B*) Corresponding single-MT travel distributions. Blue lines, best fits to a single exponential decay. Hatched bars at 16 μm indicate cumulative counts of travel distance that exceeded our field of view. Mean travel distance ($d \pm$ standard error), goodness of fit ($R_{adj}^2$), and sample size (*n* trajectories) are indicated.

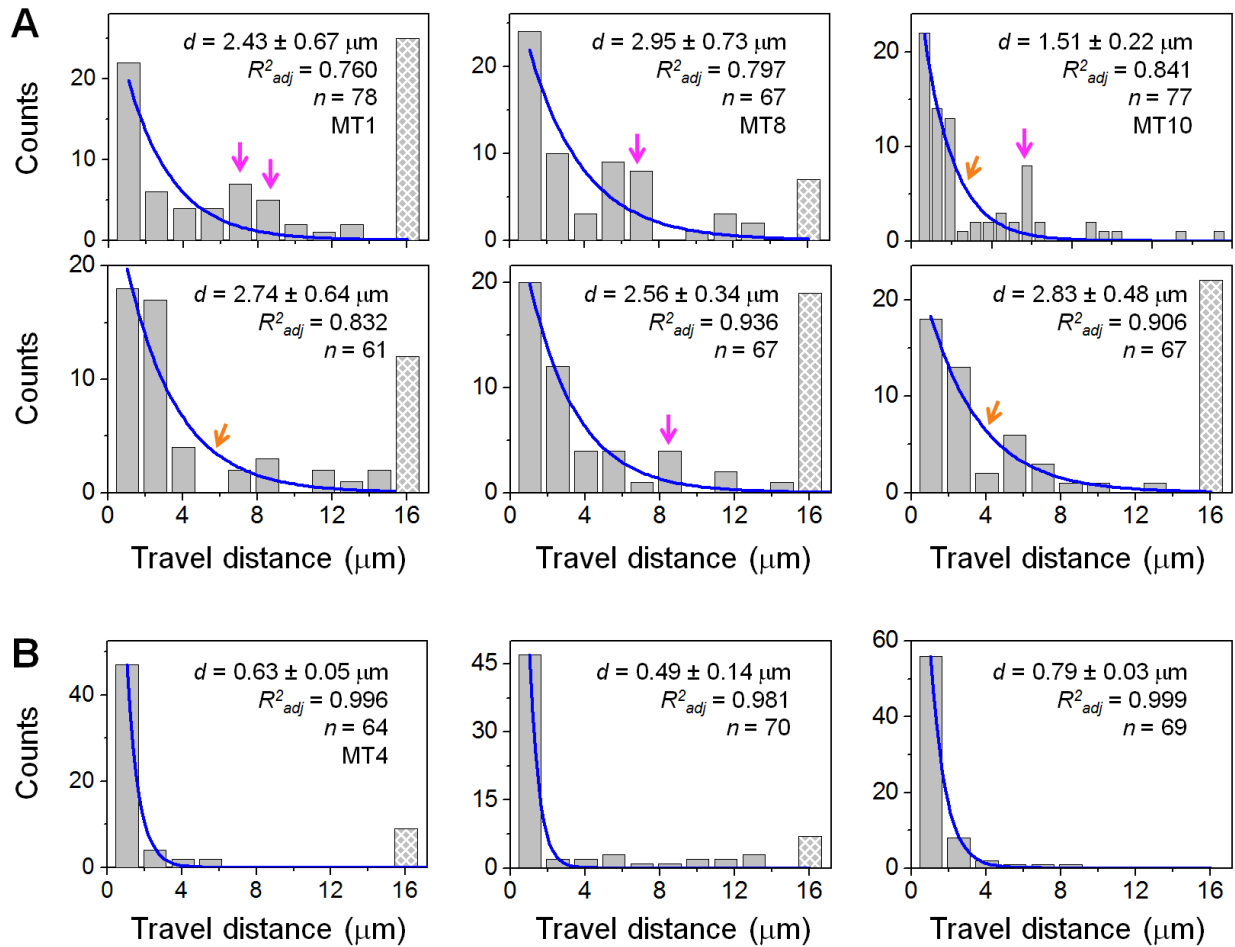

**Figure S4.** Distributions of single-microtubule (MT) travel distances, measured for few-motor transport along taxol-stabilized MTs. Nine of 33 distributions measured (27.3%) suggest that structural details in MTs have the potential to prompt kinesins to disengage from transport. Blue lines, best fits to a single exponential decay. Hatched bars at 16 μm indicate cumulative counts of travel distances that exceeded our field of view. Mean travel distance ($d \pm$ standard error), goodness of fit ($R^2_{adj}$), and sample size ($n$ trajectories) are indicated. MTs are numbered as in Figure 1 in the main text. The kinesin-to-bead ratio was kept constant at 1.4 nM kinesins:$2.8\times10^6$ beads/μL (few-motor range, Fig. S2). (*A*) Six single-MT travel distributions exhibiting unusual distinctions from a typical single exponential decay. Arrows, deviations from best fit (magenta, more counts; orange, fewer counts; see Materials and Methods). (*B*) Three single-MT travel distributions with mean travel distance substantially smaller (>3.5 standard error) than the single-kinesin travel distance (12).

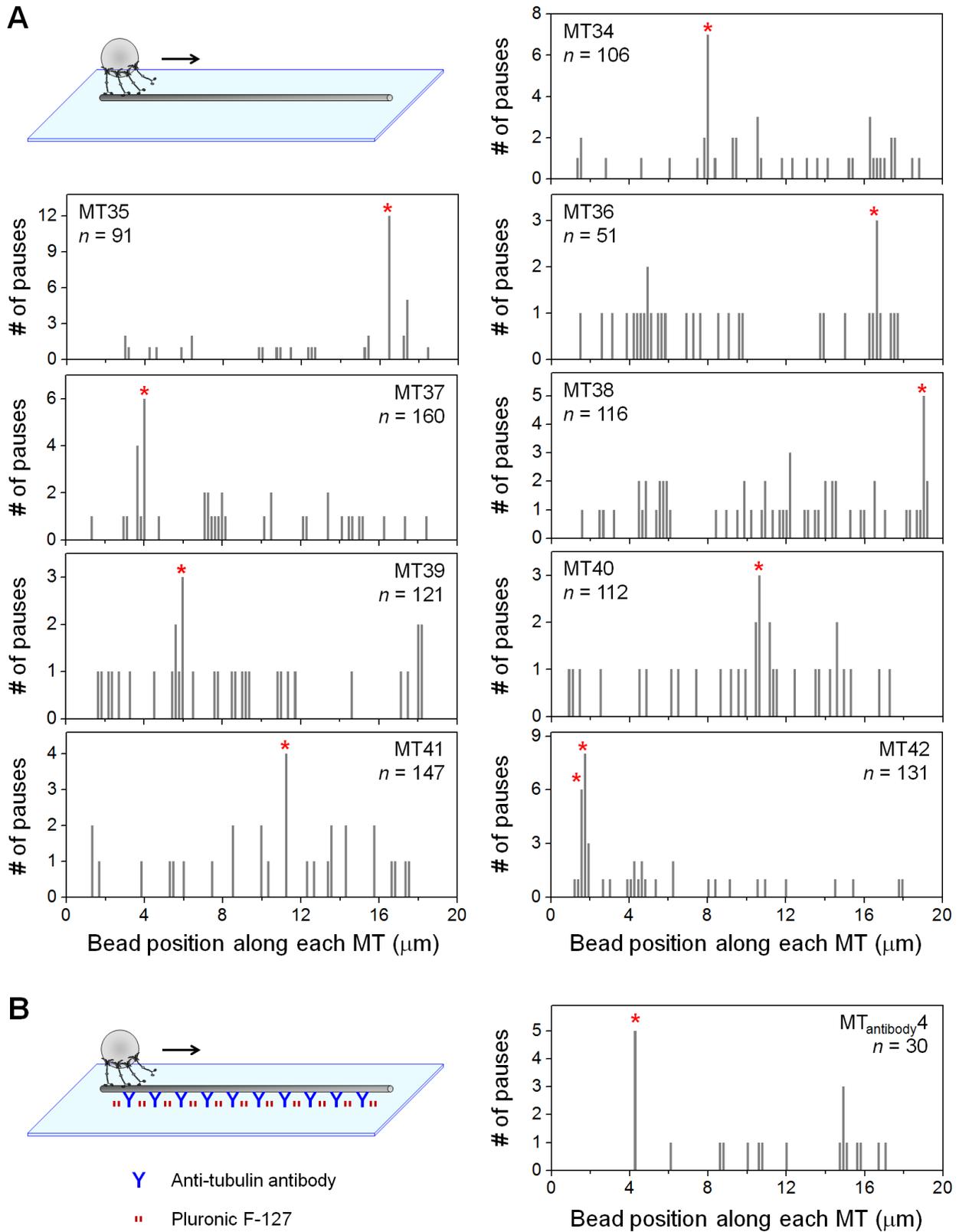

**Figure S5.** Distributions of pauses during many-motor transport along individual taxol-stabilized microtubules (MTs). Experimental schematic (not to scale), the location of common pause

locations (red asterisks, >4 standard deviations above the mean), and the number of trajectories measured for each MT (*n*) are indicated. (*A*) Measurements of cargo pausing along nine polylysine-supported MTs. (*B*) Measurements of cargo pausing along an antibody-supported MT. Of the 30 trajectories measured, we detected 5 trajectories pausing at the same site on the antibody-supported MT (~4.3 µm, asterisk). This pause count was >4 standard deviations above the mean for the MT.

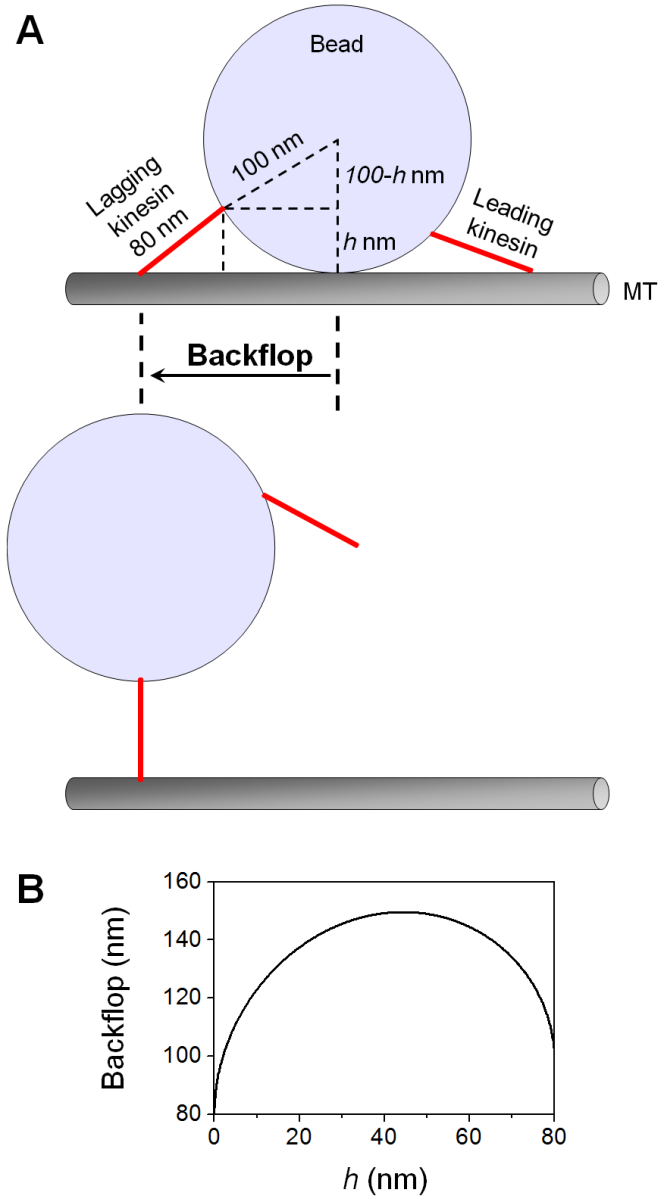

**Figure S6.** Estimated size of cargo back-flop, given kinesin's contour length (80 nm (13, 14)), our bead size (100 nm radius), and non-specific motor/bead attachment geometry ($h$ nm). (*A*) Schematic of cargo flopping back to the position of the lagging motor after the leading motor unbinds from the microtubule. Illustration is to scale. Red lines indicate kinesin motors. $h$ is determined by the relative positions of motors binding the cargo; $h$ ranges between 0 and 80 nm. (*B*) Estimated size of cargo back-flop, as determined by $\sqrt{80^2 - h^2} + \sqrt{h(200 - h)}$. The backward excursion size may be larger than estimated here, since the lagging motor may extend beyond its contour length under forward load (15).

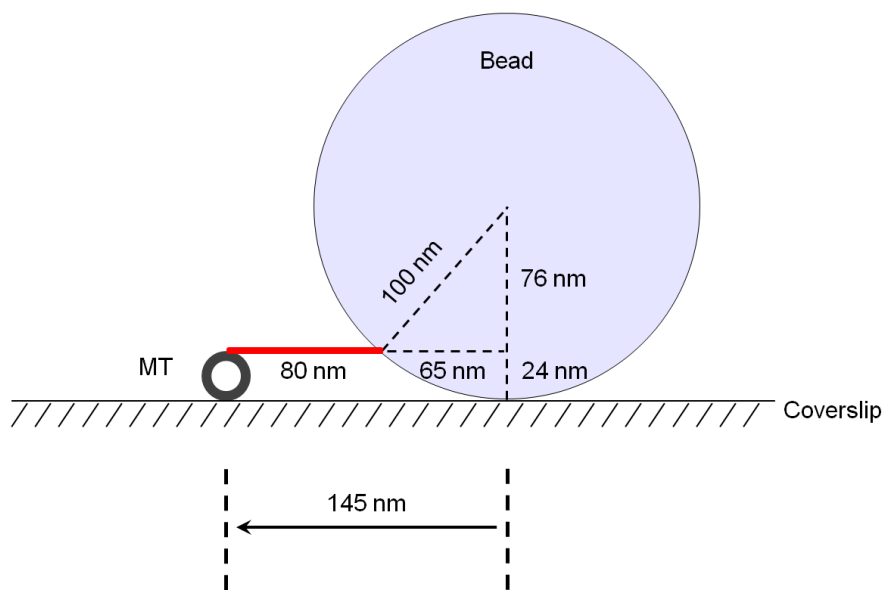

**Figure S7.** Possible range of off-axis cargo positions, given kinesin's contour length (80 nm (13, 14)) and our bead size (100 nm radius). Illustration is to scale. Red, kinesin. The range of off-axis positions observed in our study (~200 nm, Fig. 5 in the main text) is within this estimated range (290 nm) and is thus reasonable.

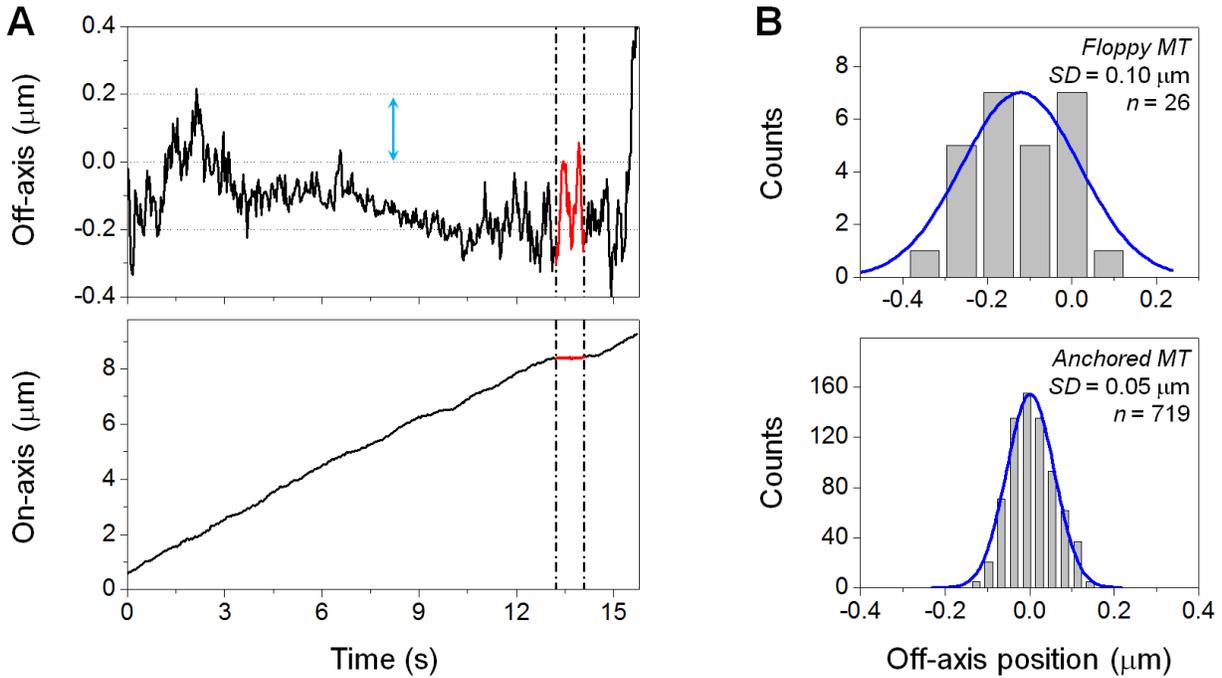

**Figure S8.** Cargo pausing on a section of a microtubule (MT; taxol-stabilized) that is not well anchored to the coverslip surface (coated with poly-L-lysine) (see also Movie S1). (*A*) Off-axis (top) and on-axis (bottom) positions for the bead trajectory. Note that the mean off-axis position is inaccurate because this MT was floppy and underwent thermal motion with respect to the coverslip. Cyan arrow and horizontal grid lines in the transverse trajectory (top) indicate the typical off-axis range of trajectories (~0.2 μm) when the MT is stably fixed to the coverslip surface (e.g., Fig. 5 in the main text). Vertical dash-dot lines indicate pausing. (*B*) Distributions of the off-axis positions of the bead during pausing for the trajectory shown in *A* (top) and for the entire trajectory on a well-anchored MT (bottom). Blue lines, best fits to a Gaussian distribution. These two distributions differ significantly ($P = 8\times10^{-47}$, two-sample *t* test). The doubling of standard deviation during pausing (0.1 μm, top; 0.05 μm, bottom) indicates that the section of MT on which pausing occurred was free from the coverslip surface.

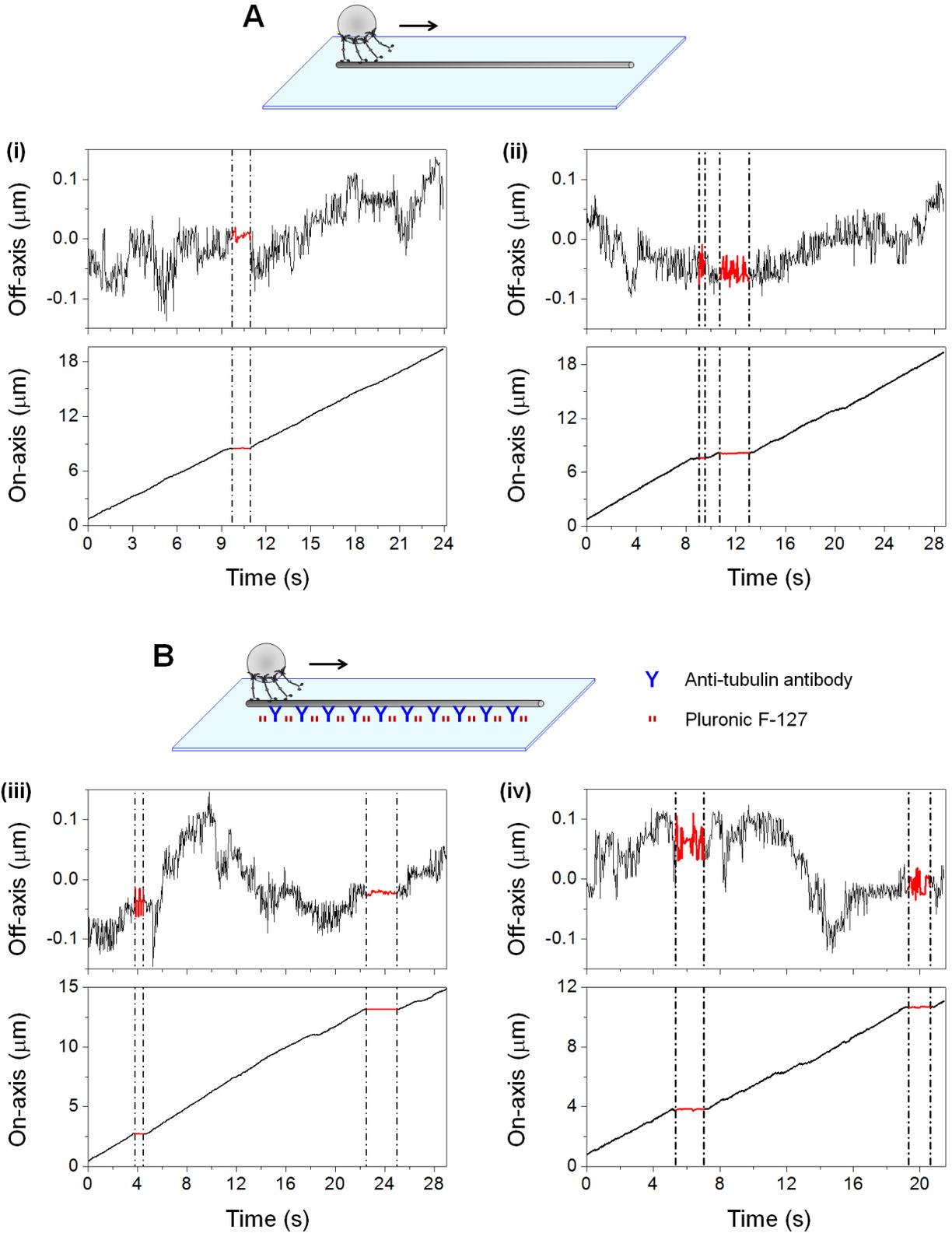

**Figure S9.** Off-axis (top) and on-axis (bottom) trajectories of beads along taxol-stabilized microtubules. Vertical dash-dot lines indicate pausing. (*A*) Measurements using polylysine-

supported microtubules. The pause in trajectory (i) corresponds to Pause 6 in Figure 5 *B* in the main text. The second pause (at ~12 s) in trajectory (ii) corresponds to Pause 2 in Figure 5 *B* in the main text. The constrained bead diffusion during pausing in (i) was not a general observation: off-axis diffusion of the bead during pausing was not constrained in (ii). The constrained off-axis diffusion during pausing in (i) did not arise from interactions between the bead and the coverslip surface, since the bead did not pause at the microtubule/coverslip interface (the off-axis position was ~0 μm). The constrained off-axis diffusion in (i) is unlikely to arise from non-specific interactions between the bead and the microtubule surface, since we previously verified that the beads used in our study require motors to interact with microtubules (16). However, we do not rule out the possibility that transient nonspecific interactions between the bead and the microtubule may account for some pauses. (*B*) Measurements using antibody-supported microtubules. The combination of antibody attachment and coating the coverslip surface with block copolymers (Pluronic F-127) reduces the possibility of a surface effect. We again observed instances of constrained off-axis diffusion during pausing (at ~24 s in iii). Again, this constrained off-axis diffusion during pausing was not a general observation (at ~4 s in iii, and both pauses in iv).

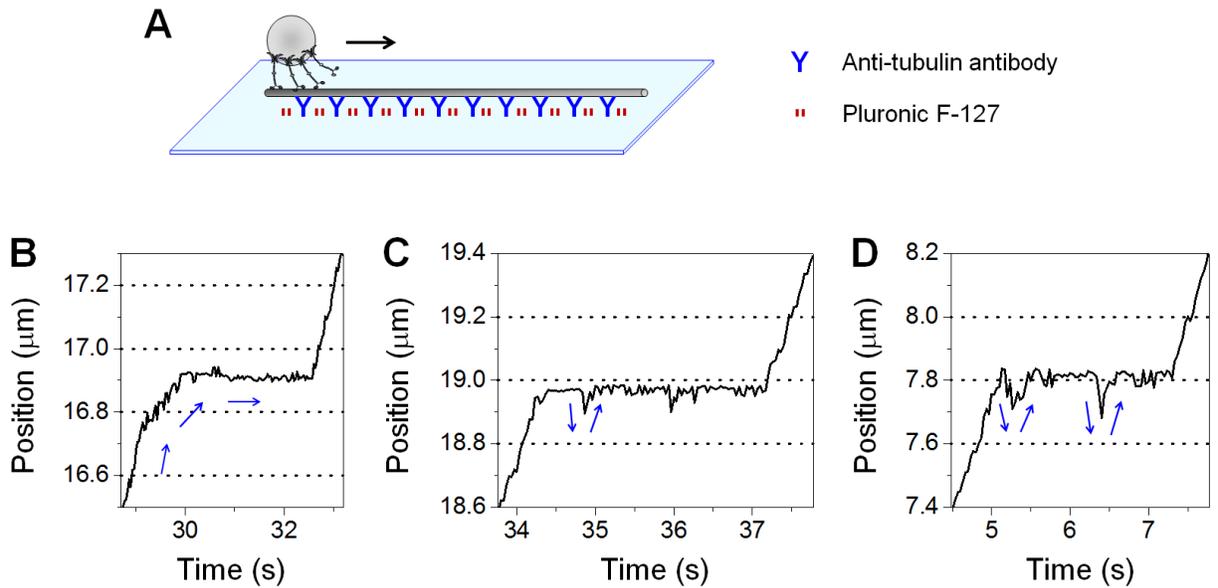

**Figure S10.** Example trajectories of cargo pausing along antibody-supported taxol-stabilized microtubules. (*A*) Experimental schematic (not to scale). (*B-D*) Example trajectories exhibiting static (*B*) and dynamic (*C-D*) pausing. Blue arrows indicate the direction of cargo travel. During dynamic pausing (*C-D*), the range of backward movements is consistent with the estimated size of cargo back-flop (Fig. S6); mean cargo velocity (± standard error) was 1.9 ± 0.3 μm/s (*n* = 4) during backward movement and 0.62 ± 0.08 μm/s (*n* = 4) during forward movement. These velocities are in excellent agreement with those measured using polylysine-supported microtubules (Fig. 4 *D-E* in the main text).

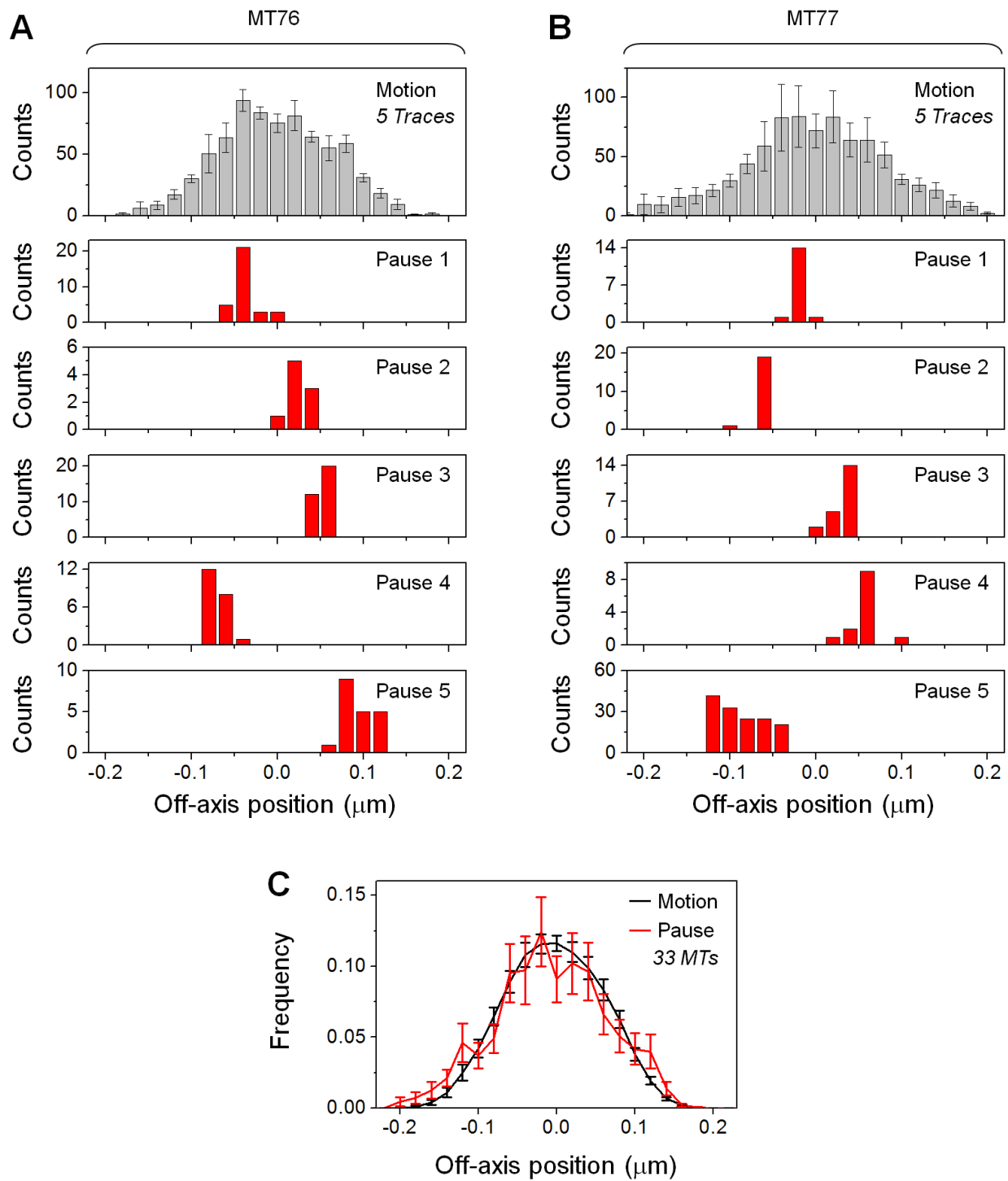

**Figure S11.** Distributions of off-axis positions of beads during pausing and during motion for taxol-polymerized microtubules (MTs). (*A-B*) Example distributions. Distributions during cargo

motion represent the averages of five trajectories each. Error bars, standard error. Distributions during cargo pausing were not averaged and are representative of individual trajectories. (*C*) Normalized distributions averaged over 33 taxol-polymerized MTs. Error bars, standard error. The distribution of off-axis positions during pausing was in excellent agreement with that during motion. Both distributions were well described by a normal distribution ($R_{adj}^2$ = 0.95 and 0.99, respectively) that is centered about the mean off-axis position of the MT (-5.9 ± 3.2 nm and -3.0 ± 1.1 nm, respectively).

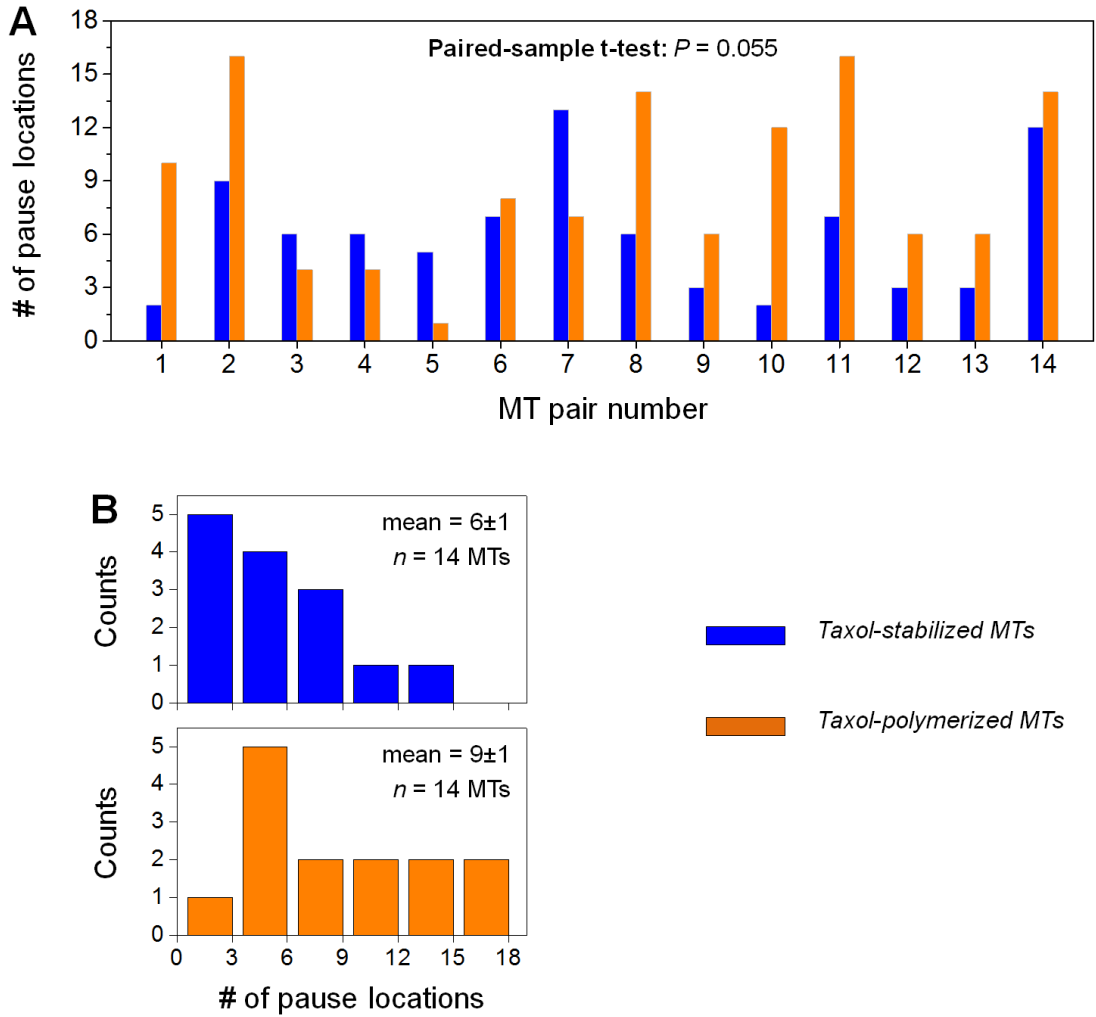

**Figure S12.** Number of pause locations on microtubules (MTs) with different defect frequencies, corresponding to pausing measurements in Figure 6 in the main text. (*A*) Number of distinct pause locations on each MT. (*B*) Distributions of the number of pause locations for each MT type. Mean number of pause locations per MT (± standard error) and sample size (*n* MTs) are indicated.

**SUPPORTING MOVIE**

**Movie S1.** Cargo pausing on a microtubule that was partially anchored to the coverslip surface. Imaging was performed with differential interference contrast microscopy and video was recorded at 30 Hz. The field of view in each frame is 9.7 μm x 2.1 μm. The corresponding trajectory appears in Figure S8 *A*. Pausing occurred between 13.2 s and 14.1 s, on a section of the microtubule that was free from the coverslip surface (Fig. S8).